
\magnification 1200

\centerline{\bf Gravity - Balls }

\centerline{}

\centerline{}
\centerline{}
\centerline{Daksh Lohiya}
\centerline{}
\centerline{\it Inter University Centre for Astromony and Astrophysics}
\centerline{\it [IUCAA], Postbag 4, Ganeshkhind}
\centerline{\it Poona, INDIA}

\vskip 2cm

\centerline{\bf Abstract}

 	 The existence of non trivial, non topological solutions in a class
of induced, effective gravity models arising out of a non minimally coupled
scalar
field is established. We shall call such solutions ``Gravity Balls'' as the
effective
gravitational constant inside the soliton differs from its effective value
outside.

\vfil\eject

I.		The conventional Einstein - Hilbert action:
$$ S = (16\pi G)^{-1}\int d^4 x \sqrt{-g} [R - 2\Lambda] \eqno{(1.1)} $$
gives a non - renormalizable quantum field theory.  Over the last three
decades, a
considerable
effort has been made to dynamically generate an effective action of gravity
[1,2].
In many such approaches, the
gravitational constant is not treated as a
fundamental parameter but rather a property of today's state of the world.
Following Zee[2] one may consider an action:

$$ S = \int d^4 x \sqrt{-g}[\epsilon\phi^2R +	g^{\mu\nu}\phi_{,\mu}\phi_{,\nu}
- 2V(\phi) + 2L_w] \eqno{(1.2)}$$
where V is a potential minimised at some value $\phi = v, L_w$	being the
lagrangian for the other matter fields. The theory, at $\phi = v$, is
indistinguishable from Einstein's theory with the gravitational constant
$[G_N = (8\pi\epsilon v^2)^{-1}]$ for small R. For large R - as in the
early Universe, the model would account for an adiabatic variation of the
gravitational constant. A perturbation  $\phi = v + \psi$ describes a scalar
particle
$\psi$ with mass $V''(\phi = v)$. At earlier
times the scalar curvature
becomes larger, giving $\delta G/G = -2\delta v/v = 2\epsilon R/V''(v)$
$\longrightarrow$ (Hubble constant / mass of $\psi)^2$. This variation would
become
important when the age $[H^{-1}]$ of the universe were to be of the order of
the Compton
time of $\psi$. Thus in these theories, G is affected by the bulk
properties only at early times while at later times it is dominated just by the
minimum of the effective potential.

		Generalising these considerations to multi component scalar fields can lead
to interesting high temperature behaviour of the effective potential in the
theory [2,3]. It is possible to avoid restoration of symmetry at high
temperatures leading to $\phi_{MIN}^2 \longrightarrow T^2$. The
effective gravitational constant weakens as $1/T^2$ in the early universe. For
the early universe, this implies that the conformal scale factor of the
Friedman - Walker metric goes linearly with time leading
to a simple resolution
of the horizon problem in cosmology [4]: The horizon radius, related to the
integral $\int dt[R(t)]^{-1}$ diverges for the conformal scale factor
$R(t) \longrightarrow t$ as the lower limit of integration goes to zero.

						Meanwhile a different
approach was proposed by Adler[5] who considered
renormalised matter action,
$$ S_A = \int d^4 x \sqrt{-g} L$$
arising out of a bare theory containing no mass [dimensional] parameters and no
scalar fields. In terms of the trace of the stress energy tensor defined by:
$$ T^{\mu\nu} = 2\sqrt{-g} \delta/\delta g_{\mu\nu}[\sqrt{-g} L]/\sqrt{-g}
\eqno{(1.3)}$$
a calculable induced gravitational constant and similarly a cosmological
constant arise in the theory by canonical prescriptions of dimensional
transmutation in field theory [6]:

$$ [16\pi G_I]^{-1} = i Lim(n\longrightarrow 4)\int d^N x (-x^2)
<T(T^\mu_\mu(x)T^\nu_\nu(0)>/96 \eqno{(1.4)}$$

$$ \Lambda_I/G_I = 2\pi <T^\mu_\mu> \eqno{(1.5)}$$
where the traces are evaluated in the flat space limit. Adler and Zee [7] have
shown that there is nothing in the above formulation that can unambigously
fix the sign of the induced gravitational and cosmological constants.

		We shall now consider particular forms of Brans - Dicke - Zee models in
the presence of additional Einstein - Hilbert terms. Both signs for the
constants appearing with the scalar curvature and the cosmological constant
term may be considered in the following:

$$S = \int d^4 x \sqrt{-g}[\epsilon\phi^2R +	g^{\mu\nu}\phi_{,\mu}\phi_{,\nu}
- 2V(\phi) + \beta_I R + \Lambda_I + 2L_w] \eqno{(1.6)}$$
Here $L_w$ is the renormalised action of the other matter fields. Reference to
the Adler - Zee program made earlier was to show that $\beta_I$ and
$\Lambda_I$ with
either sign can arise in the theory as symmetry breaking effects in quantum
field theory. Once these constants are fixed experimentally, standard
renormalization group techniques [8] imply that they are independent of the
renormalization scale. [For the present purpose, we may consider an Einstein -
Hilbert term either generated by the Adler - Zee prescription, or consider
it as having been put in by hand].As for the scalar fields,
the theory described by
the action eqn(1.7) may arise out as an effective classical phenomenological
manifestation of composite fields from some more fundamental renormalizable
theory. The basic purpose of this article is to show that in such an effective
theory classical, non - topological solutions can exist.

II.	In all that follows, we shall consider solutions in which the field has
constant
values over the interior and exterior of a  spherically symmetric region of
radius
$\approx r_0$, separated by a thin surface over which it has a gradient. For a
scalar
theory in flat spacetime, Derrick's theorem [9,10] is an impediment to the
existence of
any stable static solutions in three or more dimensions. However,
it is possible to get around Derrick's result in curved spacetime.

	Firstly, solutions approaching a non vanishing minimum of the potential
$V(\phi)$
outside a compact spatial
region ${\cal  C}$ are not forbidden in curved spacetime as they are in flat
spacetime.
These solutions would correspond to
a non - vanishing asymptotic cosmological constant. Having lost Poincare
invariance of
flat space, the conserved energy
can only be defined as the Killing energy[11]:
$$ E = \int d^3x \sqrt{-g} T^{o\nu}\zeta_\nu$$
associated with a timelike killing vector $\zeta_\nu$, of the asymptotic
spacetime. The
contribution to this energy
from the integral over $V(\phi)$ comes only from its deviation from the
asymptotic value
evaluated over ${\cal  C}$.
Thus, for example, even if one had a potential which were positive definite
[fig I], the
soliton solution having
$\phi \approx \phi_-$ = constant in the interior and $\phi \longrightarrow
\phi_+ $ outside
${\cal  C}$, would have a
negative, bounded Killing energy. With $\phi_\pm$ reversed, the solution would
have a
positive, bounded Killing energy.

	Further, the stress tensor in the above expression follows from the metric
variation
of eqn. [1.6]:

$$ [R_{\mu\nu} - g_{\mu\nu}R/2](\epsilon\phi^2 + \beta_I) +
\epsilon[\phi^2]_{;\mu;\nu}
- \epsilon g_{\mu\nu}[\phi^2]^{;\rho}_{;\rho} + \phi_{,\nu}\phi_{,\mu} +$$
$$g_{\mu\nu}[ - \phi_{,\rho}\phi^{,\rho} + 2V(\phi) - \Lambda_I]
= T_{\mu\nu}^w \eqno{(2.2)}$$
As in the Zee model, for small R, the scalar field does not affect the
background symmetry of
spacetime. Even in
the flat spacetime limit, the $\epsilon$ - dependent derivative terms of the
field contribute
to the surface term
of a typical static solution as:
$$ \epsilon\int\nabla.[\nabla\phi^2]d^3x = 2\epsilon\int |\nabla\phi|^2d^3x +
2\epsilon\int_{\cal C}\phi\nabla\phi.ds \eqno{(2.3)}$$
The second term, for a discontinuous gradient of $\phi$ at the surface
${\cal C}$, aside from a suitable choice of
sign and magnitude of $\epsilon$, could yield a contribution of the surface
term of either
sign. [The
corresponding contribution in the Derrick result comes from a positive definite
$|\nabla\phi|^2$ only].

	Finally, one must consider the gravitational contribution to the energy. This
comes
mainly on two counts.
First is the Newtonian potential energy of the solution. If the difference
between the inner
and outer values of
the potential term [$V_\pm(\phi)$] is $\epsilon$, the energy required to
assemble a ball of
radius $r_o$ is
$-16\pi^2\epsilon^2G_{out}r_o^5/45$. The second contribution comes from the
distortion of
geometry inside the ball.
The energy density of the scalar field is:
$$ \int_o^{r_o} 4\pi r^2\epsilon dr[1 - 8\pi\epsilon G_{in}r^2/3]^{-1/2}$$
This can be exactly calculated. For the present purpose we just note its small
$|8\pi G_{in}\epsilon r_o^2|$ limit
as:
$$ [4\pi/3]r_o^3\epsilon + 16\pi^2\epsilon^2G_{in}r_o^5/15 \eqno{(2.4)}$$

	In general one could consider both signs of $\beta_I$ (and hence $G_I$
and $\Lambda_I$) as well as both signs for $\epsilon$. The effective
gravitational
constant would be given by:

$$ G_{Eff} = G_I/[1 + \epsilon G_I<\phi^2>] \eqno{(2.5)}$$
To evaluate the expression for the surface energy consider the field eqn. for
$\phi$:
$$ \phi^{,\rho}_{;\rho} + 2\epsilon R\phi - V'(\phi) = 0 \eqno{(2.6)}$$
R can be eliminated from this equation and the trace of eqn(2.4) to get an
equation of the form:

$$ \phi^{,\rho}_{;\rho} + W'(\phi) = 0 \eqno{(2.7)}$$
for $\epsilon = -1/6$, for example, $W'(\phi)$ has the form:

$$ W'(\phi) = [2\Lambda_I/3\beta_I - T^{w\alpha}_\alpha/\beta_I]\phi +
[V - \phi V'/4]2\phi/\beta_I - V'(\phi)$$

One could look for static spherically symmetric solutions to this equation for
$W(\phi)$ having the profile given in figure(I) for a suitable choice of
parameters defining V and the value of $T^\rho_\rho$, i.e. $W'(\phi)$ has zeros
at $\phi = 0$ and at $\phi = \phi_o$. For small R, static,
spherically symmetric solutions to eqn(2.7), in the thin wall approximation,
satisfy:

$$ \phi'' = -2\phi'/r + W'(\phi) \eqno{(2.8)}$$
Non trivial solutions to this equation have the following general behaviour:
$\phi$ stays close to a value $\phi_o$ minimising $W(\phi) = W(\phi_o)$ for
r going from zero to a large value $R_o$ and thereafter quickly goes over to
$\phi = 0$ over a further distance of the order of $[W''(0)]^{-1/2}$. For large
radius, the behaviour of $\phi$ at the transition is given by $\phi'' =
W'(\phi)$
which has a solution:

$$ {\it R} - r = \int_{\phi_o}^\phi d\phi[2W]^{-1/2} \eqno{(2.9)}$$
The energy of
the surface is of the order:$ E_S = 4\pi {\it R}^2\int dr[\phi'^2/2 + W] =
 4\pi {\it R}^2\int_{\phi_o}^o d\phi \sqrt{2W}$.

	To summarize, in general the energy of a large soliton would be a sum
of three terms:
$$ E = Ar_o^3 + Br_o^2 + Cr_o^5 \eqno{(2.10)}$$
The sign of $A$ [the volume term] would be determined by the sign of $V_+ -
V_-$. The sign
of $B$ [the surface term] would be determined by the $\epsilon$
and the discontinuity in the gradient of $\phi$ at the surface. The sign of
$C$ [the gravitational correction] would be determined by the gravitational
constant
$ G_{EFF} $. It is straight forward to see that stable, non trivial
solutions would arise for:
$$ (i) B > 0, A < 0, C > 0$$
$$ (ii) B< 0, A < 0, C > 0$$
$$ (iii) B < 0, A > 0, C < 0$$
$$ (iv) B < 0, A > 0, C > 0$$
The vanishing of $E$ for any value of radius $r = r_b$ is the signal of the
birth of a true
vacuum bubble [12]. If $r_b < r_o$ [the soliton size], the
soliton would be unstable to vacuum decay. Thus the solitons (ii), (iii) and
(iv) are both classically and quantum mechanically stable. These solutions have
negative total energy. As regards the solution (i), the energy of the soliton
can be positive
or negative [depending on the relative values of A and C]. For the positive
energy soliton, there
is no value for $r_b$ for which $E = 0$. This solution is thus stable against
vacuum dacay while
the negative energy soliton would be unstable. The solution has $\phi$ resting
in the false vacuum outside ${\cal C}$ where the spacetime is flat. Inside
${\cal C}, \phi$
minimises $W(\phi)$, with an effective negative cosmological, constant inside.
The false vacuum
outside the solution is stabilised against vacuum decay by gravity. [This was
first pointed out by
Coleman
and deLucia[12]].

	We expect such solutions to have interesting consequences in cosmology
and shall describe their possible uses in a separate paper.
\vskip 1cm

Acknowledgement:

		Heplful discussions with Dr. Gary Gibbons, Dr. R. J. Rivers and
Dr. M. D. Pollock are gratefully acknowledged.

\vfil\eject

\centerline{\bf References}

\item{1.}  W. Scherrer, Helv. Phys. Acta 23, 547 (1950)
 C. Brans, R. Dicke, Phys. Rev. 124, 925(1961); Phys. Rev. 125, 2163 (1962).
 F. Hoyle and J. V. Narlikar, Proc. Roy. Soc. (Lond) A282, (1964) 190
 C. Callan, S. Coleman, R. Jackiw, Ann. Phys. (NY) 59, 42  (1970)
 S. Deser, Ann. Phys. 59 (1970) 248.

\item{2.} A. Zee, Phys. Rev. Lett. 42, 417 (1979); Phys. Rev. D23, 858 (198 );
Phys.
 Rev. Lett 44, 703 (1980).

\item{3.} S. Weinberg, Phys. Rev. D9, 3357 (1974).
 R. N. Mohapatra, G. Senjanovic, Phys. Rev. Lett. 42, 1651 (1979)

\item{4.} W. Rindler, Mon. Not. Roy. Ast. Soc. 116, 663 (1956)

\item{5.} S. L. Adler, Rev. Mod. Phys. 54 (1982) 729 and references therein.

\item{6.} P. M. Stevenson, Ann. Phys. (NY), 132, 383 (1981).

\item{7.} A. Zee, in ``Unity of forces in the Universe" Vol II, Ed. A. Zee,
  P. 1062, World Scientific (1982).

\item{8.} S. Weinberg, in ``Relativity'' eds. S. W. Hawking and W. Israel
  CUP (1985)

\item{9.} See eg. R. Rajaraman, ``Solitons and instantons", Northolland, 1982.

\item{10.} G. H. Derrick, J. Math. Phys.,5, 1252 (1964)

\item{11.} S. Deser, L. F. Abbot, Nucl. Phys. B195, 76 (1982);
  S. Deser,``Energy, stability and the Cosmological constant'', Brandies Univ.
Preprint 1983.

\item{12.}S. Coleman, F. deLuccia, Phys. Rew. D21, 3305 (1980)

\vskip 2cm

Figure Captions:

     [A]  ``Figure 1".

\bye